\documentclass[prb,aps,twocolumn,showpacs,superscriptaddress,longbibliography]{revtex4-2}
\usepackage{graphicx}
\usepackage{graphicx, color}
\usepackage{hyperref}
\definecolor{link}{rgb}{0.1,0.1,0.9}
\hypersetup{colorlinks=true,linkcolor=link,citecolor=link,urlcolor=link,linktocpage}
\usepackage{epsfig}
\usepackage{amsmath}
\usepackage{amsfonts}
\usepackage{amssymb}
\usepackage{graphicx}
\usepackage{color}
\usepackage{tabularx}
\usepackage{txfonts}
\usepackage{bm}
\usepackage{dcolumn}
\usepackage{xcolor}

\begin{document}

\title{Quasi-Two-Dimensional Quantum Antiferromagnetism in the Distorted Honeycomb Compound KCuIn(PO$_4$)$_2$}

\author{S. Gayen} 
 \affiliation{Department of Physics, Bennett University, Greater Noida, 201310, India}   
 \author{S. S. Ali} 
 \affiliation{Department of Physics, Bennett University, Greater Noida, 201310, India}
 \author{V. K. Singh}
 \affiliation{Department of Physics, Indian Institute of Technology Tirupati, Tirupati, 517619, India}
\author{B. Koteswararao}
 \affiliation{Department of Physics, Indian Institute of Technology Tirupati, Tirupati, 517619, India}
\author{S. K. Panda}
\email[Corresponding author: ]{swarup.panda@bennett.edu.in}
\affiliation{Department of Physics, Bennett University, Greater Noida, 201310, India}
\begin{abstract}
 We investigate the electronic structure and magnetic properties of the distorted honeycomb lattice compound KCuIn(PO$_4$)$_2$ through a combination of experimental measurements, first-principles calculations and quantum monte-carlo simulations. Density-functional theory calculations within the GGA+U framework establishes KCuIn(PO$_4$)$_2$ as an indirect-gap insulator with Cu$^{2+}$ ($S=\tfrac{1}{2}$) local moments and finite magnetocrystalline anisotropy arising from spin–orbit coupling. A microscopic evaluation of magnetic exchange interactions using the magnetic force theorem reveals a pronounced hierarchy of couplings, with the next-nearest-neighbor interaction dominating over the nearest-neighbor exchange, while interlayer couplings remain negligible. This exchange hierarchy naturally maps the system onto weakly coupled antiferromagnetic spin chains embedded in a distorted honeycomb lattice. Motivated by the \textit{ab initio} estimated exchange interactions, we construct an effective spin-$\tfrac{1}{2}$ Hamiltonian and investigate its magnetic response using large-scale quantum Monte Carlo simulations. The calculated temperature-dependent susceptibility and field-dependent magnetization quantitatively reproduce the experimental behavior and capture key signatures of low-dimensional quantum magnetism, including a broad susceptibility maximum and a field-induced saturation at low temperatures. Our results establish KCuIn(PO$_4$)$_2$ as a quasi-two-dimensional quantum antiferromagnet composed of coupled spin chains, providing a consistent theoretical framework that links electronic structure, exchange interactions, and collective magnetic behavior.   
\end{abstract}
\maketitle
\section{Introduction}
Low-dimensional quantum magnets continue to serve as ideal platforms for exploring emergent quantum phases arising from enhanced quantum fluctuations, competing exchange couplings, and reduced dimensionality. In particular, $S=1/2$ transition-metal oxides and phosphates exhibit a wide range of unconventional magnetic ground states driven by their quasi-one-dimensional (1D) or two-dimensional (2D) structural motifs~\cite{1D}. The discovery of high-temperature superconductivity in layered cuprates highlighted the rich physics originating from strongly correlated 2D systems~\cite{Bednorz1986}. More generally, the interplay of low dimensionality, electronic correlations, and lattice geometry in cuprates and related compounds continues to inspire the search for new quantum magnetic phases. Classic realizations of $S=1/2$ Heisenberg antiferromagnetic (HAFM) chains include compounds such as CuGeO$_3$ and Sr$_2$CuO$_3$~\cite{PhysRevB.58.R2913,PhysRevB.56.3402}, where strong intrachain exchange interactions lead to pronounced quantum fluctuations. CuGeO$_3$ undergoes a spin-Peierls transition into a dimerized spin-singlet state with a finite spin gap~\cite{spin-Peierls}, whereas other low-dimensional cuprates such as Cs$_4$CuSb$_2$Cl$_{12}$ display fractionalized spinon-like excitations and signatures of a quantum-spin-liquid regime~\cite{PhysRevB.101.235107}. Even in 2D cuprates, gapless spin excitations coexist with superconductivity, underscoring the important role of magnetic fluctuations~\cite{PhysRevB.99.174515,PhysRevB.92.174525}.
\par 
Recently, distorted honeycomb-lattice cuprates have emerged as an intriguing class of frustrated $S=1/2$ magnets, where the reduced coordination number ($z=3$) and structural distortions enhance quantum fluctuations and suppress long-range order. Several copper-based oxides, including Na$_2$Cu$_2$TeO$_6$, NaCuIn(PO$_4$)$_2$ and CuAl(AsO$_4$)O, exhibit nontrivial magnetic ground states arising from such distorted honeycomb networks~\cite{PhysRevB.89.174403,PhysRevB.91.144406,PhysRevB.107.214430}. Distorted honeycomb lattices are also realized in rare-earth systems such as Yb$_2$Si$_2$O$_7$, which hosts two inequivalent nearest-neighbor interactions and demonstrates strong quantum effects~\cite{PhysRevLett.123.027201}. In some of the systems, quasi-2D $J_1$-$J_2$ network, where weak interchain couplings ($J_1$) coexist with stronger intrachain interactions ($J_2$), placing the system near a quantum critical regime with gapless magnetic excitations.
\par
\begin{figure}
\includegraphics[width=1\columnwidth]{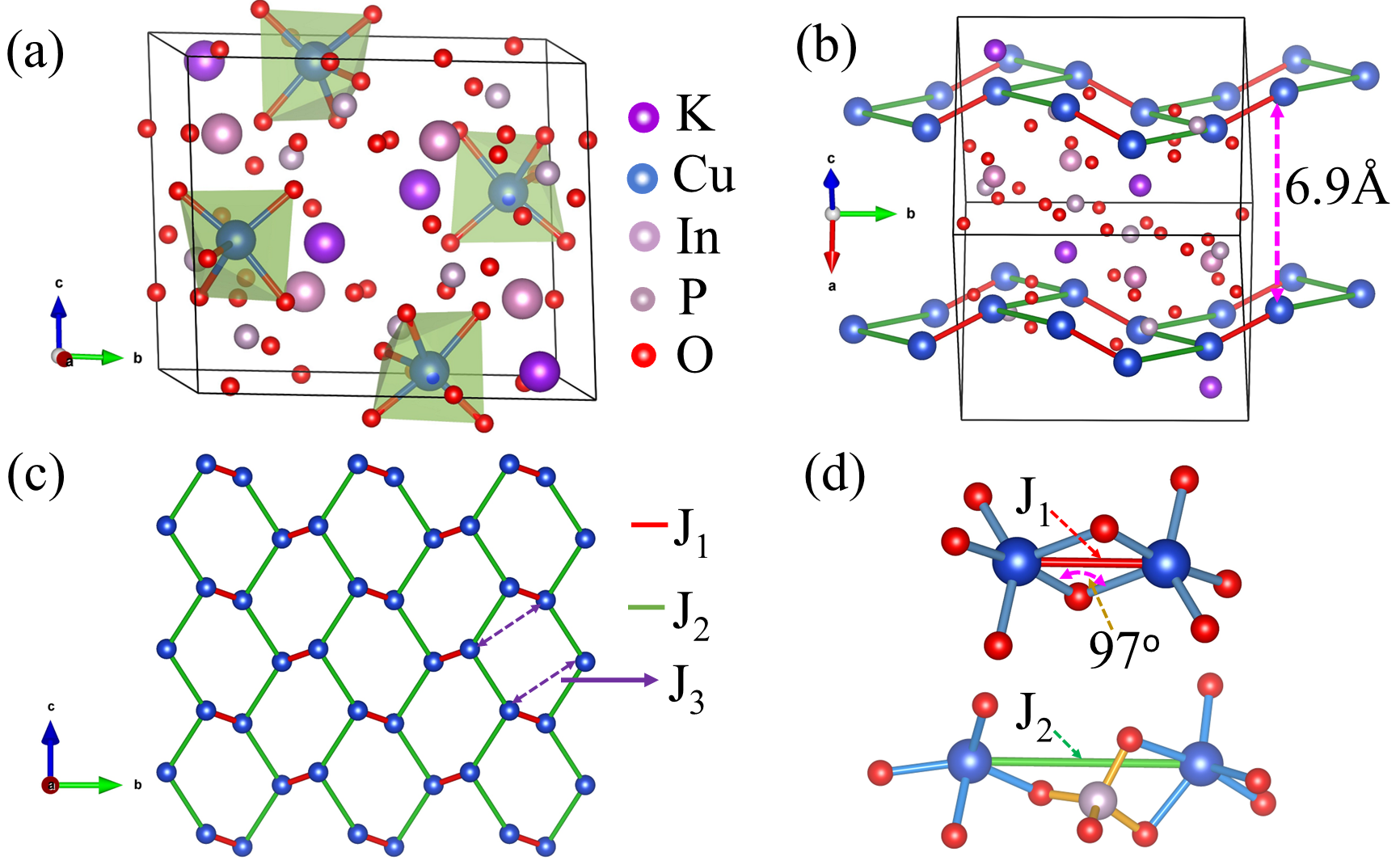}
\caption{(a) Crystal structure of KCuIn(PO$_4$)$_2$, highlighting the local environment around Cu-ions. (b) Each layer of Cu ions forming the distorted honeycomb network in the $b$-$c$ plane is displayed, and the large interlayer separation (= 6.9~\AA) between adjacent layers along the $a$-axis is highlighted, demonstrating the effectively quasi-two-dimensional nature of the magnetic lattice. (c) Each Cu-ion layer is shown separately to illustrate the formation of the distorted honeycomb geometry, with the first-, second-, and third-nearest-neighbor (NN) Cu–Cu exchange couplings labeled as $J_1$, $J_2$, and $J_3$, respectively. (d) Superexchange paths between Cu ions for the first- and second-nearest-neighbor interactions are illustrated, including the Cu–O–Cu pathway and bond angle for $J_1$, and the Cu–O–P–O–Cu pathway for $J_2$} 
\label{struct}
\end{figure}

 
In our pursuit of uncovering novel magnetic behavior in low-dimensional cuprates, we examine the newly synthesized compound KCuIn(PO$_4$)$_2$, which features $S = \frac{1}{2}$ Cu$^{2+}$ ions organized within distorted honeycomb layers. In this system, the Cu-layers are well separated along the crystallographic $a$ axis by $\sim$6.9~\AA~(see Fig.~\ref{struct}(b)), ensuring quasi-2D magnetic behavior. To elucidate the microscopic magnetic interactions in KCuIn(PO$_4$)$_2$, we performed density functional theory calculations within the GGA+$U$ framework and constructed the corresponding effective spin Hamiltonian. Our calculations suggest that within each layer, two dominant exchange pathways emerge: a nearest-neighbor (NN) intra-dimer coupling ($J_1$) with Cu-Cu distance 3.09~\AA, and a next-nearest-neighbor (NNN) intra-chain interaction ($J_2$) with Cu-Cu distance 5.34~\AA~(see Fig.~\ref{struct}(c)). The computed antiferromagnetic ground state is consistent with experimental trends and confirms the insulating nature of the compound. Motivated by the GGA+U derived exchange parameters and single-ion anisotropic energy, we carried out quantum Monte Carlo simulations. The simulated magnetic susceptibility reproduces the experimental behavior and further enables the calculation of the field-dependent magnetization, $M(H)$, at different temperatures. Our results establish KCuIn(PO$_4$)$_2$ as a quasi-two-dimensional quantum antiferromagnet composed of coupled spin chains, providing a valuable platform for exploring quantum fluctuations and spin dynamics in distorted honeycomb and $J_1$–$J_2$ magnetic systems.
\section{Experimental and Computational Methods}
\subsection{Sample Synthesis and Characterization}
Polycrystalline samples of KCuIn(PO$_4$)$_2$ were synthesized using a conventional solid-state reaction technique. High-purity precursors-CuO (Aldrich, 99.99\%), K$_2$CO$_3$ (Aldrich, 99.99\%), (NH$_4$)$_2$HPO$_4$ (Aldrich, 99.5\%), and In$_2$O$_3$ (Aldrich, 99.9\%) were weighed in stoichiometric proportions and thoroughly mixed in an agate mortar to ensure uniform homogeneity. The mixed powder was pressed into pellets and subjected to a sequence of preheating steps at 300~$^\circ$C, 400~$^\circ$C, and 500~$^\circ$C for 6~h each to promote precursor decomposition and partial reaction. A final sintering was carried out at 870~$^\circ$C for 48~h with two intermediate grinding and repelletizing steps to improve phase formation. Owing to the volatility of alkali-metal species at elevated temperatures, an additional 5\% excess of K$_2$CO$_3$ was added to compensate for potassium loss during sintering. Phase purity and structural integrity were confirmed by powder X-ray diffraction (XRD) at room temperature using a PANalytical diffractometer equipped with Cu K$\alpha$ radiation ($\lambda = 1.5406$~\AA). Rietveld refinements of the diffraction patterns were performed using the \textsc{FullProf Suite}~\cite{rodriguez1993recent}, yielding lattice parameters and atomic positions consistent with the distorted-honeycomb crystal structure. Magnetization measurements were carried out in the temperature range 2-300~K using a vibrating-sample magnetometer (VSM), with applied magnetic fields up to 7~T. These measurements provided the experimental Curie–Weiss temperature and field-dependent magnetization curves used to validate the theoretical model.
\begin{figure}
\centering
\includegraphics[width=0.85\columnwidth]{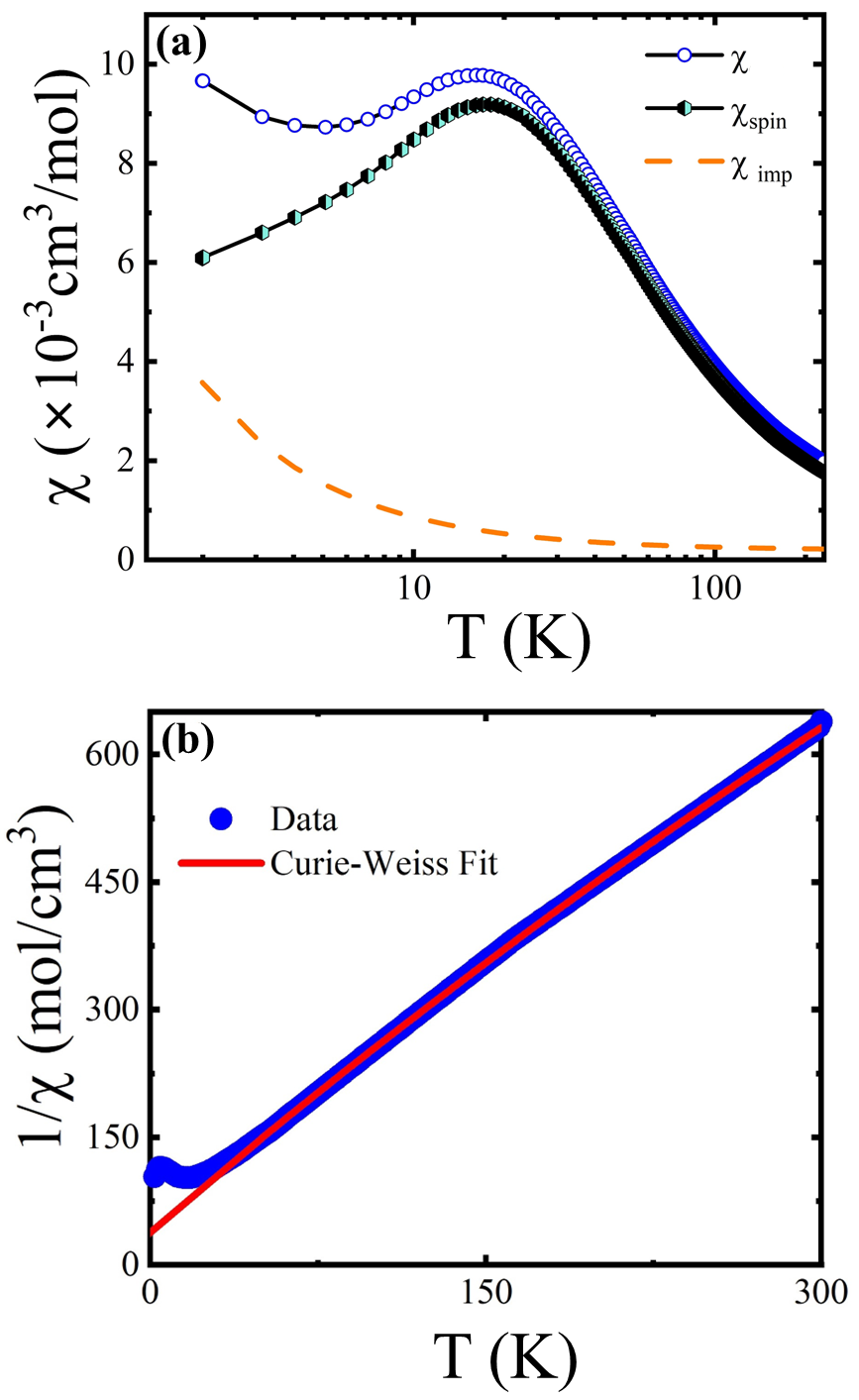}
\caption{\label{exp}(a) The temperature dependence of the magnetic susceptibility $\chi(T)$ is presented. The blue open circles represent the measured experimental data collected at an applied field of 1 T. The yellow dashed curve illustrates the impurity contribution, $\chi_{\mathrm{imp}}(T)$. The intrinsic spin susceptibility, $\chi_{\mathrm{spin}}(T)$, is extracted by subtracting $\chi_{\mathrm{imp}}(T)$ from the total measured $\chi(T)$. (b) The inverse susceptibility $1/\chi$ is plotted as a function of temperature $T$. The solid red line represents the Curie-Weiss fit, highlighting the linear behavior in the high-temperature regime.}    
\end{figure}
\subsection{Electronic-Structure Calculations}
The electronic and magnetic properties of KCuIn(PO$_4$)$_2$ were examined using density functional theory (DFT)~\cite{DFT1,DFT2}. Plane-wave calculations were performed within the VASP package~\cite{vasp1,vasp2} using the projector-augmented-wave (PAW) method~\cite{paw}. Exchange–correlation effects were treated within the generalized gradient approximation supplemented with an on-site Hubbard correction (GGA+$U$)~\cite{gga}. To capture the localized nature of Cu-$3d$ electrons, we used Hubbard's $U = 6$~eV and Hund’s coupling $J = 0.8$~eV, parameters commonly adopted for Cu-based insulating oxides~\cite{Cu-U1,Cu-U2,Cu-U3,SG1}. To assess the robustness of our conclusions against the choice of correlation strength, we also performed additional calculations for $U = 4$~eV and $U = 8$~eV. The Brillouin zone was sampled using a $7\times 5\times 6$ Monkhorst–Pack mesh and a kinetic-energy cutoff of 500~eV. The total energy convergence has been verified with respect to the plane-wave cutoff to ensure the reliability of the calculated electronic and magnetic properties. Spin–orbit coupling effects were incorporated within the framework of noncollinear density functional theory.
\par 
To extract microscopic magnetic interactions, additional full-potential linear muffin-tin orbital (FP-LMTO) calculations were performed using the RSPt code~\cite{rspt1}. After self-consistent convergence within the GGA+$U$ framework, the intersite magnetic exchange parameters $J_{ij}$ were computed using the magnetic force theorem~\cite{mft1,mft2}. This approach has been widely used for transition-metal magnetic systems~\cite{TM1,TM2,TM3}. The resulting dominant interactions define the effective magnetic model for the distorted honeycomb lattice. Magnetocrystalline anisotropic energy (MAE) is also calculated from GGA + $U$ + spin-orbit coupling approach. Using the exchange parameters obtained from the FP-LMTO calculations, we constructed an effective spin-$\tfrac{1}{2}$ Heisenberg model to describe the magnetism in KCuIn(PO$_4$)$_2$. The model incorporates the dominant isotropic exchange interactions within the distorted honeycomb lattice formed by the Cu$^{2+}$ ions. In addition, a single-ion anisotropy term was included to account for the uniaxial magnetic anisotropy revealed by the spin–orbit coupling calculations, with the preferred spin alignment along the crystallographic $z$ axis. Within this framework, the local magnetic moments are treated as quantum spins residing on the Cu$^{2+}$ sites of the lattice. The magnetic ground state and thermodynamic properties of this model were investigated using large-scale quantum Monte Carlo (QMC) simulations. We employed the stochastic series expansion (SSE)~\cite{sandvik2019stochasticseriesexpansionmethods} method with loop and directed-loop updates as implemented in the Algorithms and Libraries for Physics Simulations (ALPS) package~\cite{Bauer_2011}, which provides unbiased and numerically exact results for quantum spin systems. Simulations were performed on lattices containing up to 1600 spins with periodic boundary conditions. Extensive equilibration and sampling ensured convergence for each temperature and field value. From these simulations, we calculated magnetic susceptibility $\chi(T)$ and magnetization $M(H)$ as a function of applied field ($H$).
\par
We emphasize that all electronic-structure and magnetic calculations are performed using the experimentally refined lattice parameters and atomic positions reported in the structural characterization section of the manuscript. As the crystal structure was determined experimentally, no additional structural optimization was carried out in the present study.
\section{\label{sec:results} Results and Discussion}
\subsection{Experimental analysis}
\subsubsection{Structural characterization}
X-ray diffraction (XRD) was used to verify the structural quality and phase purity of the synthesized KCuIn(PO$_4$)$_2$ sample. The measured diffraction pattern shows excellent agreement with the simulated pattern obtained from the corresponding crystallographic information file (CIF), confirming the formation of a single-phase compound. Rietveld refinement yields reliability factors of $\chi^{2} \approx 3.25$, $R_{p} \approx 10.4\%$, $R_{wp} \approx 10.6\%$, and $R_{\mathrm{exp}} \approx 6.28\%$, demonstrating good refinement quality. The extracted lattice parameters: $a = 8.196$~\AA, $b = 10.169$~\AA, $c = 9.173$~\AA, and $\beta = 115.682^{\circ}$, are consistent with previous structural reports~\cite{badri2023synthesis}. The obtained space group is monoclinic ($P2_1/n$), and the overall crystal structure is depicted in Fig.~\ref{struct}. The lattice is composed of a network of corner- and edge-sharing polyhedra: PO$_4$ tetrahedra, CuO$_5$ triangular bipyramids, InO$_6$ octahedra, and distorted KO$_5$ coordination units. A key structural feature governing the magnetic behavior lies in the arrangement of the CuO$_5$ units. Two adjacent CuO$_5$ polyhedra share an edge, creating a nearest-neighbor (NN) Cu-Cu dimer path, denoted as $J_1$ in Fig.~\ref{struct}(b,c). The next-nearest-neighbor interaction $J_2$ connects Cu atoms through the intervening PO$_4$ tetrahedral groups, linking the $J_1$ dimer chains into an extended network [Fig.~\ref{struct}(b,c)]. This Cu-O-P-O-Cu superexchange pathway [Fig.~\ref{struct}(d)] spans a relatively long Cu-Cu distance of approximately 5.34~\AA. Although geometrically longer, this interaction plays a crucial role in establishing the dominant intrachain coupling within the distorted honeycomb lattice. Together, the 1$^{st}$ ($J_1$) and 2$^{nd}$ NN ($J_2$) generate a two-dimensional, yet distorted, honeycomb arrangement of Cu$^{2+}$ ($S=\tfrac{1}{2}$) spins. The separation between adjacent honeycomb layers is about 6.98~\AA, underscoring the quasi-two-dimensional character of the magnetic lattice and supporting the relevance of a 2D spin model for describing the magnetism of KCuIn(PO$_4$)$_2$.
\subsubsection{Magnetization Measurements}
In order to understand magnetism, we have measured the temperature dependence of the magnetic susceptibility $\chi(T)$ of polycrystalline KCuIn(PO$_4$)$_2$ as shown in Fig.~\ref{exp}. Since measured susceptibility also include contributions from dilute paramagnetic impurities, the total susceptibility can be written as
\begin{equation}
\chi(T) = \chi_{0} + \frac{C_{\mathrm{imp}}}{T + \theta_{\mathrm{imp}}} + \chi_{\mathrm{spin}}(T),
\end{equation}
where the second term models a small Curie-like upturn at low temperatures and $\chi_{\mathrm{spin}}(T)$ is the intrinsic contribution of the Cu-based spin lattice. The intrinsic susceptibility extracted after subtracting the impurity and temperature-independent contributions is displayed in Fig.~\ref{exp}(a). The measured data (blue open symbols) collected at 1~T reveal the characteristic broad peak, while the impurity contribution (yellow dashed line) accounts for the low-temperature deviation. The resulting $\chi_{\mathrm{spin}}(T)$ captures the intrinsic magnetic response of the distorted honeycomb network. A broad maximum appears around 18~K, a characteristic signature of low-dimensional $S=\frac{1}{2}$ antiferromagnets where short-range correlations develop well above any long-range ordering temperature. Such a feature typically reflects the onset of strong quantum fluctuations associated with quasi-1D or quasi-2D magnetic lattices. To quantify the high-temperature behavior, the susceptibility data in the range 20-300~K were fitted using the Curie–Weiss (CW) expression,
\begin{equation}
    \chi(T) = \chi_{0} + \frac{C}{T - \theta_{\mathrm{CW}}},
\end{equation}
where $\chi_0$ represents the temperature-independent susceptibility, $C$ is the Curie constant, and $\theta_{\mathrm{CW}}$ is the Curie–Weiss temperature. The fit shown in Fig.~\ref{exp}(b) yields 
$\chi_{0} \approx -1.38 \times 10^{-4}\,\mathrm{cm^3/mol}, 
C \approx 0.43\,\mathrm{cm^3\,K/mol},$
$\theta_{\mathrm{CW}} \approx -17~\mathrm{K}.$
The negative $\theta_{\mathrm{CW}}$ confirms dominant antiferromagnetic interactions among the Cu$^{2+}$ ($S=\tfrac{1}{2}$) moments. The temperature-independent term $\chi_0$ contains contributions from core diamagnetism and Van Vleck paramagnetism. Summing the diamagnetic terms of the constituent ions (K$^{+}$, Cu$^{2+}$, P$^{5+}$, O$^{2-}$) gives $\chi_{\mathrm{dia}} \approx -1.50\times 10^{-4}$\,cm$^3$/mol~\cite{selwood2013magnetochemistry}. Subtracting this from $\chi_0$ yields a small Van Vleck component, $\chi_{\mathrm{VV}} \approx 1.2\times 10^{-5}$\,cm$^3$/mol, arising from second-order mixing of excited states in an external magnetic field. From the Curie constant, the effective magnetic moment is obtained using 
$\mu_{\mathrm{eff}} = \sqrt{\frac{3k_B C}{N_A}} = \sqrt{8C}\,\mu_B \approx 1.85\,\mu_B.$
This value is slightly larger than the spin-only expectation of 1.73~$\mu_B$ for an $S=\tfrac{1}{2}$ ion with $g=2$. Such an enhancement is commonly observed in Cu$^{2+}$ systems and is attributed to residual spin–orbit coupling and slight deviations from perfect orbital quenching~\cite{PhysRevB.107.214430,PhysRevB.89.014407}.
\begin{figure}
    \centering
    \includegraphics[width=1\linewidth]{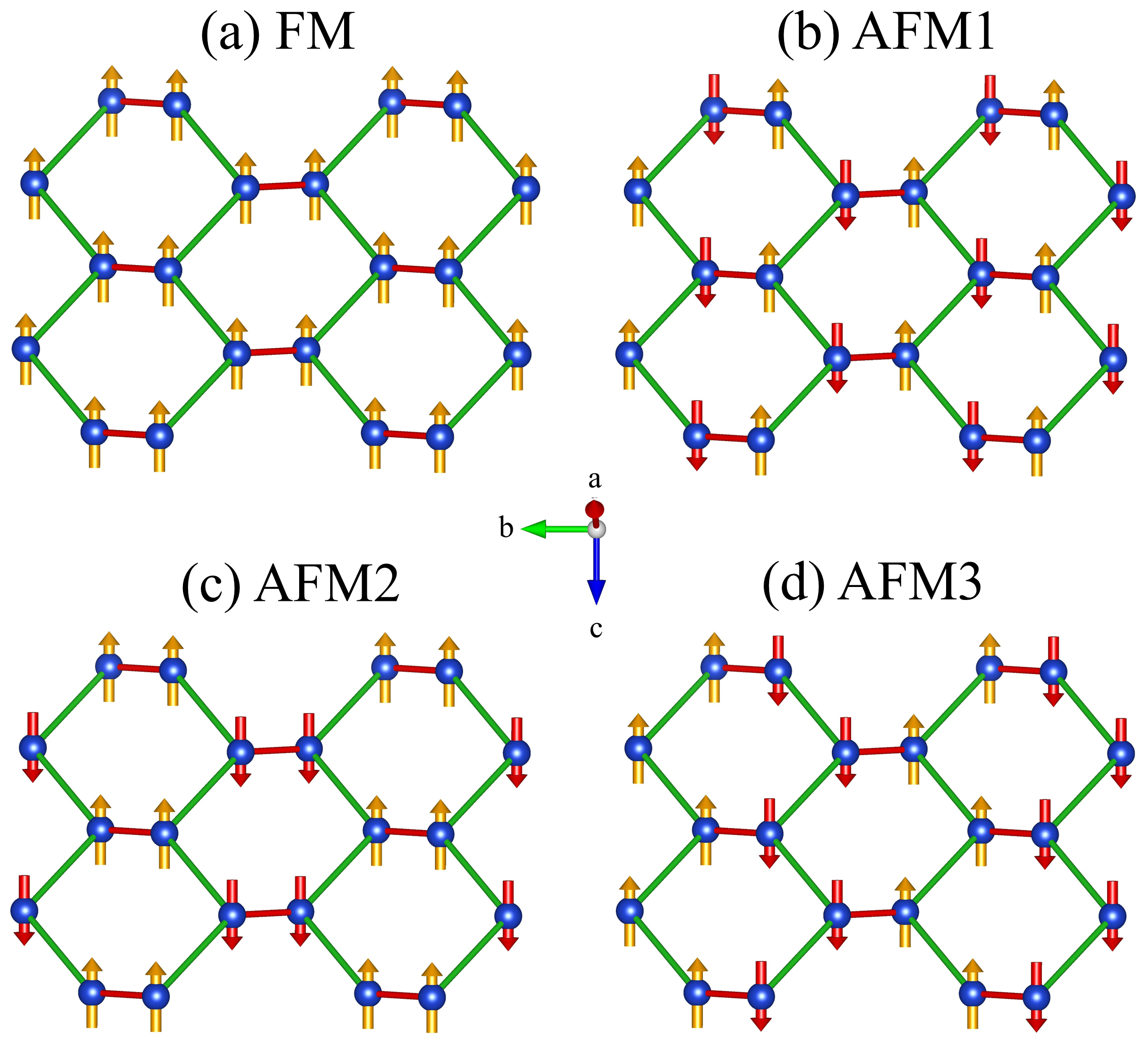}
    \caption{Four distinct Cu$^{2+}$ spin configurations within the distorted honeycomb plane are shown: (a) a ferromagnetic state with all spins aligned parallel, (b) both first- and second-nearest-neighbor interactions are antiferromagnetic, (c) the first-nearest-neighbor interaction is ferromagnetic while the second-nearest-neighbor interaction is antiferromagnetic, and (d) the first-nearest-neighbor interaction is antiferromagnetic while the second-nearest-neighbor interaction is ferromagnetic.}
    \label{config}
\end{figure}
\subsection{Theoretical analysis}

\subsubsection{Lowest energy state}
To complement the experimental magnetization results and to gain microscopic insight into the magnetic interactions governing KCuIn(PO$_4$)$_2$, we carried out a systematic first-principles analysis of the electronic structure and magnetism of this system. As a first step, we evaluated the relative stability of several plausible magnetic configurations within the crystallographic unit cell. Four distinct spin arrangements of the Cu$^{2+}$ ions in the hexagonal plane, namely one ferromagnetic (FM) and three antiferromagnetic (AFM) arrangements were constructed, as illustrated in Fig.~\ref{config}.  Total energies for each configuration were calculated within the GGA+$U$ framework, and the results are summarized in Table~\ref{energy}. Among the tested spin arrangements, the AFM1 configuration exhibits the lowest energy, lying 4.76~meV per formula unit below the FM state. This finding establishes that both the nearest-neighbor ($J_1$) and next-nearest-neighbor ($J_2$) Cu-Cu interactions favor antiferromagnetic alignment. The AFM3 configuration is found to be the next energetically favorable state, whereas AFM2, in which $J_1$ is ferromagnetic and $J_2$ antiferromagnetic, is significantly less stable. These energetic trends reflect the hierarchy of magnetic couplings and reinforce the experimentally observed negative Curie-Weiss temperature ($\theta_{\mathrm{CW}} \approx -17$~K). The identification of AFM1 as the lowest energy state provides a strong basis for the subsequent theoretical analysis. We also examined the robustness of our results with respect to the choice of correlation strength. Thus, we carried out additional calculations for $U$=4 eV and $U$=8 eV. In both cases, the AFM1 configuration remains the energetically preferred state, being lower than the FM state by 5.53 meV/f.u. (for $U$= 4 eV) and 2.51 meV/f.u. (for $U$ = 8 eV). These results demonstrate that the dominant magnetic interactions remain antiferromagnetic and the overall nature of the inter-atomic magnetic couplings is insensitive to moderate variations of Hubbard $U$ within a physically reasonable range. In later sections, we further quantified the exchange parameters for $U$ = 6 eV to determine the relevant spin model for this system and solve it using quantum Monte Carlo simulations. This establishes a coherent connection between experiment, electronic structure, and model Hamiltonian approaches.
\begin{table}[htbp]
\centering
\caption{Total energies (meV/formula unit) of the antiferromagnetic configurations shown in Fig.~\ref{config}, referenced to the ferromagnetic state, obtained within the GGA+$U$ framework.}
\vspace{2mm}
\setlength{\tabcolsep}{15pt}
\begin{tabular}{ c  c  c  c }
\hline\hline
\multicolumn{4}{c}{Energy (meV/f.u.)} \\[1mm]
\hline
  FM \hspace{0.02cm} & \hspace{0.02cm} AFM1 \hspace{0.02cm} & \hspace{0.02cm}  AFM2 \hspace{0.02cm} & \hspace{0.02cm}  AFM3 \hspace{0.02cm} \\
\hline
0.00 & -4.76 & -0.30 & -2.12 \\
\hline\hline
\end{tabular}
\label{energy}
\end{table}

\begin{figure}
\includegraphics[width=1\columnwidth]{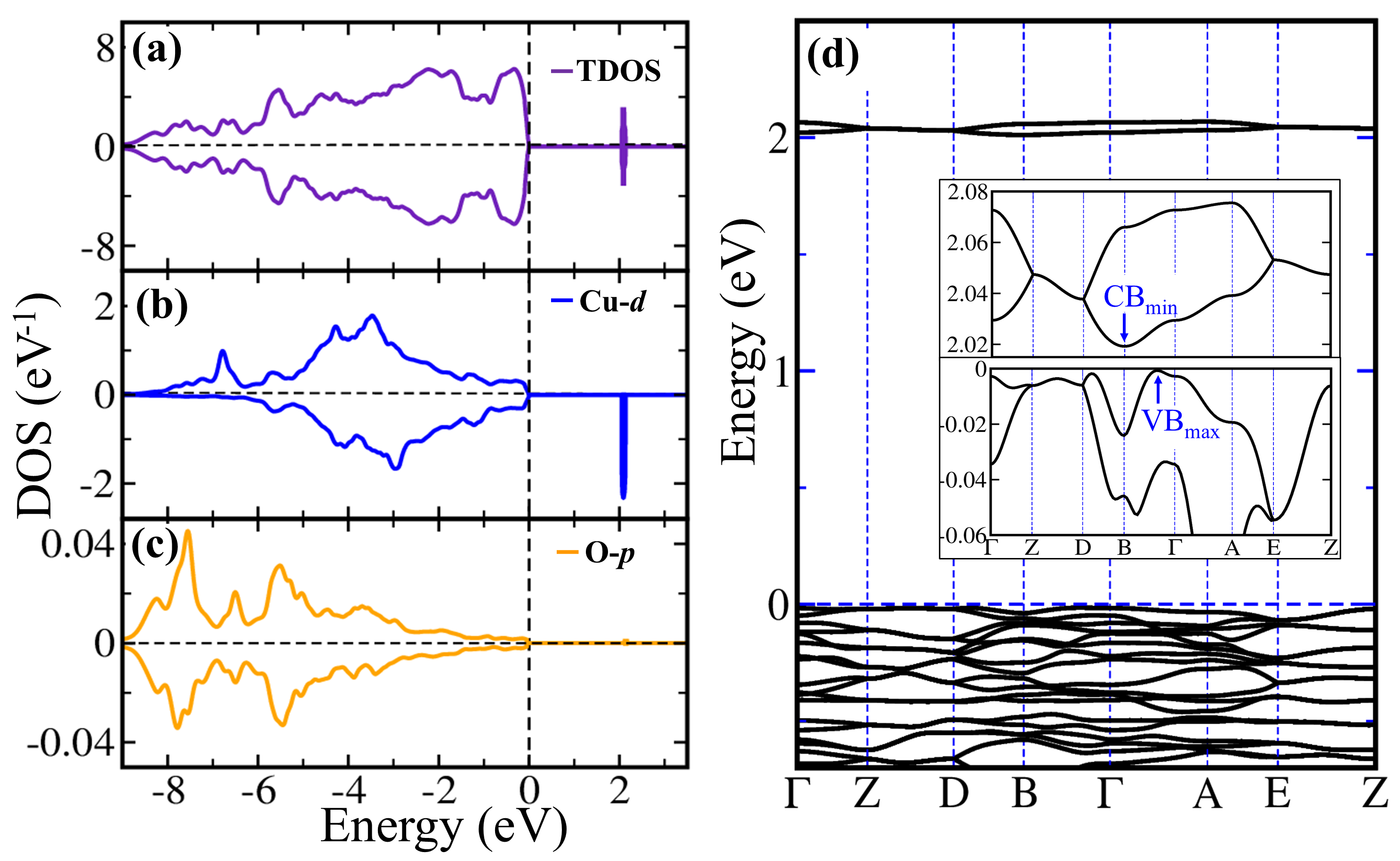}
\caption{(a) Total electronic density of states (DOS) of KCuIn(PO$_4$)$_2$}. (b) Partial density of states corresponding to the Cu–$d$ orbitals and (c) the O–$p$ orbitals. (d) Electronic band dispersion plotted along the high-symmetry directions of the Brillouin zone. The inset highlights the positions of the valence-band maximum and conduction-band minimum, illustrating the indirect-gap nature of the insulating ground state. The Fermi energy is set at 0 eV for all the figures.
\label{dos-band}
\end{figure}
\subsubsection{Electronic structure}
Before analyzing the magnetic interactions in detail, it is essential to understand the underlying electronic structure and the spin state of the Cu ions, which fundamentally determine the magnetic behavior of KCuIn(PO$_4$)$_2$. Thus, we computed the total and projected density of states (DOS) for the Cu-$d$ and O-$p$ orbitals, as shown in Fig.~\ref{dos-band}. The total DOS confirms that the compound is an insulator, exhibiting a band gap of approximately 2.0~eV. We note that this value represents a theoretical prediction based on the GGA+$U$ framework, and direct experimental verification of the band gap is not presently available. A closer examination of the projected DOS provides further insight into the orbital character near the Fermi level. As illustrated in Fig.~\ref{dos-band}(a), the majority-spin Cu-$d$ states are fully occupied in the valence band, while the minority-spin states remain partially filled. The unoccupied minority-spin states appear in the conduction band, consistent with the expected $d^9$ electronic configuration of Cu$^{2+}$, in which one unpaired electron contributes to the local magnetic moment. The calculated magnetic moment on the Cu site is 0.73~$\mu_{\mathrm{B}}$, slightly lower than the ideal ionic value of 1.0~$\mu_{\mathrm{B}}$ for an isolated Cu$^{2+}$ ion. This reduction is commonly attributed to hybridization between Cu-$d$ and O-$p$ orbitals. We note here that computed magnetic moment corresponds to the static local moment within the Cu atomic sphere as obtained from spin-polarized DFT+U calculations. It cannot be directly compared with the effective paramagnetic moment obtained from Curie–Weiss analysis,  presented in the experimental section. The paramagnetic moment is a thermodynamic quantity extracted in the high-temperature paramagnetic regime and characterizes the response of fluctuating local moments. In general, DFT-derived local moments are most appropriately compared with ordered moments measured by neutron diffraction, where the static spin density is directly probed. To date, no neutron measurements are available for this material. We also present the electronic band dispersion along the high-symmetry \textit{k}-paths of the monoclinic Brillouin zone in Fig.~\ref{dos-band}(d). To elucidate the nature of the band gap, the region near the Fermi level is highlighted in the inset, focusing on the top of the valence band and the bottom of the conduction band. This analysis reveals that KCuIn(PO$_4$)$_2$ possesses an indirect band gap where the conduction-band minimum is located at the B point, while the valence-band maximum occurs along the B–$\Gamma$ direction of the Brillouin zone.
\par 
Given the known influence of spin–orbit coupling (SOC) in Cu$^{2+}$ systems, we explicitly included SOC in our calculations to quantify its magnitude and implications for magnetic anisotropy. The GGA+U+SOC calculations yield an orbital moment of 0.16~$\mu_{\mathrm{B}}$ per Cu, resulting in an orbital-to-spin ratio of approximately 0.22. This ratio indicates a significant effective SOC.  To evaluate the magnetocrystalline anisotropy, we computed the total energy for magnetization constrained along the crystallographic $a$, $b$, and $c$ axes. The system exhibits uniaxial anisotropy with the $c$ axis as the easy axis, lower in energy by 0.26~meV/f.u. compared to the $a$ axis. 
This analysis therefore establishes both the insulating ground state and the expected $S=\tfrac{1}{2}$ nature of the Cu$^{2+}$ ions with non-negligible SOC-induced anisotropy. These electronic features form an essential microscopic basis for understanding the magnetic interactions and low-dimensional behavior discussed in the subsequent sections.
\begin{table}[t]
\centering
\caption{Magnetic exchange interactions between Cu–Cu neighbors evaluated using the magnetic force theorem. These couplings corresponding to the distinct Cu–Cu neighbor pairs are labeled as indicated in Fig.~\ref{struct}.}
\vspace{2mm}
\setlength{\tabcolsep}{8pt} 
\begin{tabular}{c c c c c}
\hline\hline
$J_{1}$ (meV) & $J_{2}$ (meV) & $J_{3}$ (meV) & $J_{2}/J_{1}$ & $J_{3}/J_{1}$ \\
\hline
$-0.90$ & $-4.35$ & $-0.15$ & 4.83 & 0.16 \\
\hline\hline
\end{tabular}
\label{exchange}
\end{table}
\subsubsection{Interatomic Exchange Couplings}
Having established the preferred spin configuration within the crystallographic unit cell, we next turned to evaluating the interatomic magnetic exchange interactions ($J_{ij}$) that govern the emergent magnetic properties of KCuIn(PO$_4$)$_2$. A quantitative determination of these couplings is crucial for constructing an accurate spin Hamiltonian and for understanding the low-dimensional magnetic behavior suggested by the susceptibility and structural analyses. To compute the exchange parameters, we employed the magnetic force theorem~\cite{mft1,mft2} as implemented in the \textsc{RSPt} code. The resulting exchange constants are summarized in Table~\ref{exchange}. All significant couplings are antiferromagnetic, consistent with the negative Curie–Weiss temperature obtained experimentally. A striking feature of the calculated exchanges is that the next-nearest-neighbor (NNN) interaction $J_2$ is substantially larger than the nearest-neighbor (NN) interaction $J_1$, with a ratio $J_2/J_1 \approx 4.8$. In contrast, the interlayer coupling $J_3$ is extremely weak, with $J_3/J_1 \approx 0.16$, underscoring the quasi-two-dimensional nature of the magnetic lattice. The small magnitude of $J_1$ can be traced to the local bonding geometry. The Cu–O–Cu bond angle mediating $J_1$ is $97.34^\circ$ [Fig.~\ref{struct}(d)], close to the $90^\circ$ limit where the Goodenough–Kanamori rules~\cite{GEK} predict weak ferromagnetic superexchange. In this geometry, the ferromagnetic superexchange interaction is likely to compete with the direct antiferromagnetic exchange arising from the overlap of half-filled Cu $d$ orbitals. This situation originates from the nominal $d^9$ electronic configuration of Cu$^{2+}$ ions, for which the ligand field enforces a single half-filled $d$ orbital. The competition between these two contributions can lead to partial cancellation, resulting in a reduced net exchange strength and in the present case, the effective exchange interaction comes out to be antiferromagnetic. In contrast, the dominant $J_2$ interaction couples Cu ions through the extended Cu–O–P–O–Cu superexchange pathway mediated by the PO$_4$ tetrahedra [Figs.~\ref{struct}(b,c)]. Despite the relatively long Cu–Cu separation (5.34~\AA), this NNN path allows for efficient superexchange due to favorable orbital overlaps and a near-linear O–P–O linkage. To elucidate why the next-nearest-neighbor interaction $J_{2}$ is stronger than the nearest-neighbor interaction $J_{1}$, we analyzed the Cu–Cu hopping amplitudes using a Wannier-based tight-binding model. The model was constructed from maximally localized Cu $d_{x^{2}-y^{2}}$–like orbitals, which constitute the dominant orbital character near the Fermi level. The Wannier interpolation accurately reproduces the DFT band structure, enabling the extraction of the dominant hopping parameters. The nearest-neighbor hopping is $t_{1} = 39.10$~meV, while the NNN hopping is substantially larger, $t_{2} = 72.90$~meV. In contrast, the interlayer hopping is much smaller, $t_{3} = 20$~meV. Since the magnetic superexchange in a half-filled Cu$^{2+}$ system scales as $J \sim 4t^{2}/U$, the ratios of the exchange interactions can be estimated directly from the hopping amplitudes. We obtain
$\frac{t_{2}^{2}}{t_{1}^{2}} = 3.6$, which closely matches the magnetic-force-theorem estimate $J_{2}/J_{1} \approx 4.8$. This clearly shows that the larger NNN hopping is responsible for the dominant $J_{2}$ interaction, despite the longer Cu-Cu separation. Similarly, the small ratio $t_{3}^{2}/t_{1}^{2} \approx 0.26$ is consistent with the negligible interlayer exchange $J_{3}$. Thus, both hopping-based estimates and total-energy-derived exchange interactions yield the hierarchy $J_{2} > J_{1} \gg J_{3}$, demonstrating that KCuIn(PO$_4$)$_2$ effectively realizes a quasi-two-dimensional distorted honeycomb antiferromagnet, governed primarily by strong intrachain antiferromagnetic $J_2$ interactions, with weaker dimerlike $J_1$ links and negligible interlayer coupling. This forms the basis for the quantum Monte Carlo analysis presented in the following section.
\begin{figure}[htbp]
\includegraphics[width=1\columnwidth]{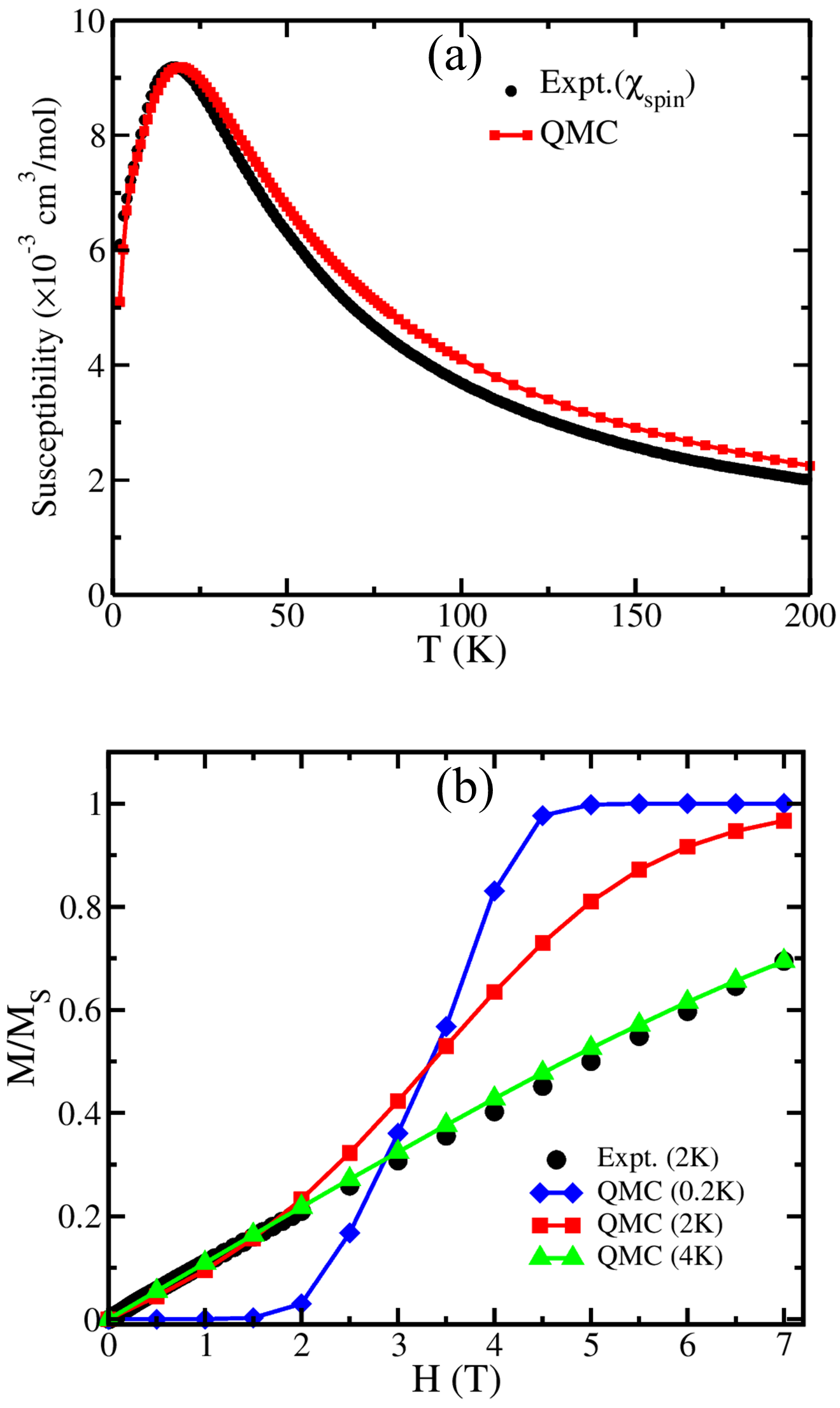}
\caption{(a) Temperature-dependent magnetic susceptibility calculated using quantum Monte Carlo simulations for the effective spin-$\tfrac{1}{2}$ Hamiltonian (red squares). The simulated susceptibility shows good agreement with the experimental susceptibility, $\chi_{\mathrm{spin}}(T)$ (black circles). (b) Field-dependent normalized magnetization, $M/M_s$ at different temperatures, compared with the corresponding experimental magnetization curves. The calculations capture the characteristic low-temperature saturation behavior and its evolution with temperature, demonstrating the consistency of the effective spin model with the experimental observations.}
\label{qmc}
\end{figure}
\subsubsection{Quantum monte-carlo simulations}
The microscopic exchange analysis based on first-principles calculations establishes a clear hierarchy of magnetic interactions in KCuIn(PO$_4$)$_2$, with the next-nearest-neighbor coupling $J_2$ significantly larger than the nearest-neighbor interaction $J_1$. This hierarchy naturally suggests a magnetic picture dominated by quasi-one-dimensional antiferromagnetic chains formed by $J_2$, with weaker but non-negligible interchain coupling mediated by $J_1$. Guided by this insight, we performed large-scale quantum Monte Carlo (QMC) simulations to quantitatively assess whether such a spin Hamiltonian can reproduce the experimentally observed magnetic response. We constructed the following effective spin-$\frac{1}{2}$ Heisenberg Hamiltonian. 
\begin{equation}
H = -J_1 \sum_{\langle i,j \rangle} \hat{S}_i\cdot \hat{S}_j 
-J_2 \sum_{\langle\langle i,j \rangle\rangle} \hat{S}_i\cdot \hat{S}_j - \sum_i K({{S}^z_i})^2,
\end{equation}
where $J_1$ corresponds to the nearest-neighbor (intra-dimer) interaction and $J_2$ represents the next-nearest-neighbor (intrachain) coupling within the distorted honeycomb network. $\hat{S_i}$ and $\hat{S_j}$ denote the unit vector of local spin moment of Cu$^{2+}$ ion at $i^{th}$ and $j^{th}$ site of the lattice. $K$ symbolize for single-ion anisotropy constant  and ${S}^z_i$ is the spin component along the preferred crystallographic \text{z}-axis. Using the above effective spin-$\tfrac{1}{2}$ Heisenberg model, we carried out QMC simulations within the stochastic series expansion framework. We find that a parameter set with $J_2 = 30$~K and $J_1 = 8$~K reproduces the experimental susceptibility remarkably well, yielding a ratio $J_2/J_1 \approx 3.75$. This ratio is in close agreement with the independently obtained estimate from the magnetic force theorem within the GGA+$U$ approach, thereby providing strong internal consistency between electronic-structure calculations, model Hamiltonian analysis, and experimental observations. Figure~\ref{qmc}(a) compares the intrinsic experimental susceptibility $\chi_{\mathrm{spin}}(T)$ with the QMC results. The simulations successfully capture the broad maximum around $T \sim 18$~K, a hallmark of low-dimensional antiferromagnets dominated by short-range correlations, as well as the overall temperature dependence over a wide range ($0.5$-$200$~K). This agreement confirms that the magnetism of KCuIn(PO$_4$)$_2$ is well described by a minimal spin Hamiltonian consisting of dominant $J_2$ chains weakly coupled by $J_1$, together with a uniaxial anisotropy term. Further insight is obtained from the field-dependent magnetization. Figure~\ref{qmc}(b) shows the magnetization isotherms obtained from QMC simulations at $T = 0.2$, $2$, and $4$~K, along with the experimental data measured at $T = 2$~K. The simulated $M(H)$ curve at $4$~K closely follows the experimental data, exhibiting an approximately linear field dependence and the absence of saturation up to the highest measured field of 7~T. 
This behavior is characteristic of antiferromagnetic spin-chain systems, where strong exchange interactions suppress full spin polarization at accessible fields. We observe that the experimental M(H) data measured at 2 K agree very well with the theoretically simulated curve at 4 K, while differences is present between experiment and the theoretical 2 K result. Such discrepancies at very low temperatures can arise from minor uncertainties in the exchange parameters and the finite statistical resolution inherent to stochastic QMC sampling. In this regime, thermal broadening is minimal, and even small variations in the DFT-derived coupling constants can lead to such differences in the magnetization slope. Since the experimental measurements are limited to $T= 2$~K, we further employed QMC simulations to explore the low-temperature regime inaccessible experimentally. At $T = 0.2$~K, the magnetization curve exhibits a clear saturation at a magnetic field of approximately 5~T. Such field-induced saturation reflects the closure of the spin gap and the transition from a singlet-dominated ground state to a fully polarized triplet state, a phenomenon widely observed in low-dimensional $S=\tfrac{1}{2}$ antiferromagnets under strong magnetic fields. The emergence of saturation at relatively moderate fields further supports the dominance of one-dimensional physics governed by $J_2$, with the weaker $J_1$ coupling providing interchain connectivity without destroying the essential spin-chain character. These results firmly establish KCuIn(PO$_4$)$_2$ as a quasi-one-dimensional quantum antiferromagnet with a distorted honeycomb topology.
\section{\label{sec:conclusion} Conclusion} 
In summary, we have presented a comprehensive theoretical study of the electronic and magnetic properties of KCuIn(PO$_4$)$_2$, combining first-principles electronic-structure calculations with quantum Monte Carlo simulations to establish a consistent microscopic description of its low-dimensional magnetism. Our first-principles electronic-structure calculations reveals that the compound is an indirect-gap insulator with well-localized Cu$^{2+}$ ($S=\tfrac{1}{2}$) moments and finite magnetocrystalline anisotropy, arsing due to the presence of spin–orbit-coupling. A detailed evaluation of magnetic exchange interactions uncovers a pronounced hierarchy of couplings, with the next-nearest-neighbor interaction $J_2$ dominating over the nearest-neighbor coupling $J_1$, while interlayer interactions are negligible. This exchange topology places KCuIn(PO$_4$)$_2$ in a regime of weakly coupled antiferromagnetic spin chains embedded in a distorted honeycomb lattice, leading to an effective quasi-two-dimensional magnetic network. Based on the \textit{ab initio} exchange parameters, an effective spin-$\tfrac{1}{2}$ Hamiltonian accurately reproduces the experimentally observed magnetic susceptibility and magnetization through large-scale quantum Monte Carlo simulations. The broad maximum in $\chi(T)$ and the nearly linear magnetization response at finite temperature are characteristic of strong short-range antiferromagnetic correlations and gapless spin excitations expected in low-dimensional quantum magnets dominated by chain-like interactions.
\par 
A central outcome of this work is the identification of the ratio $J_2/J_1$ as the key control parameter governing the magnetic ground state of KCuIn(PO$_4$)$_2$. When $J_2 \gg J_1$, the system approaches the physics of nearly isolated Heisenberg spin chains, characterized by strong quantum fluctuations and gapless excitations. Reducing this ratio is expected to enhance two-dimensional correlations and eventually stabilize a N\'eel-ordered state, as observed in related distorted honeycomb and coupled-chain systems. Conversely, increasing $J_2/J_1$ further would push the material deeper into the spin-chain regime. Such a continuous evolution between N\'eel order and one-dimensional quantum magnetism has been extensively discussed in theoretical studies of $J_1$-$J_2$ models and observed experimentally in several Cu-based low-dimensional magnets. From a materials perspective, the pronounced sensitivity of the exchange hierarchy to local bonding geometry suggests that external perturbations such as uniaxial strain, pressure, or chemical substitution could provide viable routes to tune the ratio $J_2/J_1$ in KCuIn(PO$_4$)$_2$. Modifying Cu-O-Cu bond angles or Cu-O-P-O--Cu superexchange pathways may allow controlled access to different magnetic regimes, including proximity to quantum critical points separating ordered and disordered phases. In this sense, KCuIn(PO$_4$)$_2$ emerges as a promising platform for exploring tunable low-dimensional quantum magnetism in distorted honeycomb lattices.
\par 
Overall, our study establishes KCuIn(PO$_4$)$_2$ as a model system in which electronic structure, exchange interactions, and collective magnetic behavior are tightly linked. The combined \textit{ab initio} and many-body approach presented here provides a robust framework for understanding and engineering quantum magnetism in a broader class of low-dimensional cuprate-based materials.
\section{ACKNOWLEDGMENTS}
S.K.P. and  B.K. acknowledges support from ANRF (previously
SERB), Government of India for the core research grant
(CRG/2023/003063). The computations were enabled by resources provided by Bennett University.
\section{APPENDIX: Stability analysis}
We assess the dynamical stability of the experimentally determined crystal structure by performing first-principles $\Gamma$-point phonon calculations using the finite-displacement method~\cite{PhysRevLett.78.4063} as implemented in VASP code~\cite{vasp1,vasp2}. The phonon frequencies were obtained by diagonalizing the dynamical matrix derived from the calculated force constants. As shown in Table~\ref{phonon}, all 156 phonon modes coming from 52 atoms in the unit cell (3N modes, N being the total number of atoms in a unit cell) exhibit positive frequencies, with the three acoustic modes approaching zero as expected. The absence of imaginary (negative) modes confirms the dynamical stability of the structure. The phase purity and structural integrity is already established by X-ray diffraction measurements. Hence, both experimental and theoretical analyses consistently demonstrate that the crystal structure is stable. and the presented calculations are performed using the experimentally refined structure.
\begin{table*}[!h]
\centering
\renewcommand{\arraystretch}{1.2}
\caption{Calculated $\Gamma$-point phonon modes and their corresponding frequencies (in THz) obtained using the finite-displacement method.}
\begin{tabular}{c c @{\hspace{1cm}} c c @{\hspace{1cm}} c c @{\hspace{1cm}} c c @{\hspace{1cm}} c c @{\hspace{1cm}} c c}
\hline
Mode & {Freq.} & Mode & {Freq.} & Mode & {Freq.} &  Mode & {Freq.} & Mode & {Freq.} & Mode & {Freq.}  \\
\hline
1   & 0.231  & 27  & 3.620  & 53  & 5.747  & 79  & 9.590  & 105 & 16.276 & 131 & 28.088 \\
2   & 0.245  & 28  & 3.648  & 54  & 5.775  & 80  & 9.830  & 106 & 16.294 & 132 & 28.151 \\
3   & 0.760  & 29  & 3.735  & 55  & 5.788  & 81  & 10.081 & 107 & 16.519 & 133 & 28.231 \\
4   & 1.315  & 30  & 3.826  & 56  & 5.905  & 82  & 10.223 & 108 & 16.533 & 134 & 28.360 \\
5   & 1.508  & 31  & 3.826  & 57  & 5.941  & 83  & 10.342 & 109 & 16.632 & 135 & 28.822 \\
6   & 1.630  & 32  & 3.863  & 58  & 5.969  & 84  & 10.361 & 110 & 16.645 & 136 & 29.002 \\
7   & 1.799  & 33  & 3.933  & 59  & 6.189  & 85  & 10.924 & 111 & 16.778 & 137 & 30.408 \\
8   & 1.878  & 34  & 3.948  & 60  & 6.349  & 86  & 11.190 & 112 & 16.835 & 138 & 30.550 \\
9   & 2.214  & 35  & 3.998  & 61  & 6.550  & 87  & 11.748 & 113 & 17.251 & 139 & 30.573 \\
10  & 2.452  & 36  & 4.149  & 62  & 6.780  & 88  & 11.912 & 114 & 17.301 & 140 & 30.603 \\
11  & 2.510  & 37  & 4.310  & 63  & 7.014  & 89  & 12.232 & 115 & 17.378 & 141 & 30.853 \\
12  & 2.563  & 38  & 4.378  & 64  & 7.167  & 90  & 12.499 & 116 & 17.455 & 142 & 31.312 \\
13  & 2.621  & 39  & 4.461  & 65  & 7.198  & 91  & 12.579 & 117 & 17.612 & 143 & 31.492 \\
14  & 2.732  & 40  & 4.477  & 66  & 7.666  & 92  & 12.706 & 118 & 17.650 & 144 & 31.866 \\
15  & 2.858  & 41  & 4.558  & 67  & 7.768  & 93  & 12.978 & 119 & 17.769 & 145 & 31.868 \\
16  & 2.870  & 42  & 4.691  & 68  & 7.865  & 94  & 13.023 & 120 & 17.830 & 146 & 31.951 \\
17  & 2.991  & 43  & 4.731  & 69  & 8.561  & 95  & 13.127 & 121 & 17.895 & 147 & 32.106 \\
18  & 3.007  & 44  & 4.778  & 70  & 8.592  & 96  & 13.132 & 122 & 17.945 & 148 & 32.285 \\
19  & 3.129  & 45  & 4.911  & 71  & 8.595  & 97  & 13.789 & 123 & 17.952 & 149 & 32.509 \\
20  & 3.150  & 46  & 4.944  & 72  & 8.719  & 98  & 14.105 & 124 & 18.140 & 150 & 33.273 \\
21  & 3.252  & 47  & 5.004  & 73  & 8.953  & 99  & 14.116 & 125 & 26.941 & 151 & 33.337 \\
22  & 3.307  & 48  & 5.067  & 74  & 9.347  & 100 & 14.172 & 126 & 26.967 & 152 & 33.656 \\
23  & 3.417  & 49  & 5.285  & 75  & 9.447  & 101 & 15.358 & 127 & 27.459 & 153 & 33.801 \\
24  & 3.473  & 50  & 5.336  & 76  & 9.510  & 102 & 15.556 & 128 & 27.713 & 154 & 34.751 \\
25  & 3.492  & 51  & 5.539  & 77  & 9.517  & 103 & 15.652 & 129 & 27.984 & 155 & 34.779 \\
26  & 3.514  & 52  & 5.548  & 78  & 9.587  & 104 & 15.787 & 130 & 28.074 & 156 & 34.966 \\
\hline
\end{tabular}
\label{phonon}
\end{table*}
\bibliography{KCuIn}

@article{PhysRevB.101.235107,
  title = {Spinon excitations in the quasi-one-dimensional $S=\frac{1}{2}$ chain compound $\mathrm{C}{\mathrm{s}}_{4}\mathrm{CuS}{\mathrm{b}}_{2}\mathrm{C}{\mathrm{l}}_{12}$},
  author = {Tran, Thao T. and Pocs, Chris A. and Zhang, Yubo and Winiarski, Michal J. and Sun, Jianwei and Lee, Minhyea and McQueen, Tyrel M.},
  journal = {Phys. Rev. B},
  volume = {101},
  issue = {23},
  pages = {235107},
  numpages = {8},
  year = {2020},
  month = {Jun},
  publisher = {American Physical Society},
  doi = {10.1103/PhysRevB.101.235107},
  url = {https://link.aps.org/doi/10.1103/PhysRevB.101.235107}
}

@article{Bednorz1986,
  author    = {Bednorz, J. G. and M{\"u}ller, K. A.},
  title     = {Possible high Tc superconductivity in the Ba--La--Cu--O system},
  journal   = {Zeitschrift f{\"u}r Physik B Condensed Matter},
  year      = {1986},
  volume    = {64},
  number    = {2},
  pages     = {189--193},
  month     = jun,
  issn      = {1431-584X},
  doi       = {10.1007/BF01303701},
  url       = {https://doi.org/10.1007/BF01303701}
}

@article{PhysRevB.99.174515,
  title = {Gapless spin excitations in superconducting ${\mathrm{La}}_{2\ensuremath{-}x}{\mathrm{Ca}}_{1+x}{\mathrm{Cu}}_{2}{\mathrm{O}}_{6}$ with ${T}_{c}$ up to 55 K},
  author = {Schneeloch, John A. and Zhong, Ruidan and Stone, M. B. and Zaliznyak, I. A. and Gu, G. D. and Xu, Guangyong and Tranquada, J. M.},
  journal = {Phys. Rev. B},
  volume = {99},
  issue = {17},
  pages = {174515},
  numpages = {10},
  year = {2019},
  month = {May},
  publisher = {American Physical Society},
  doi = {10.1103/PhysRevB.99.174515},
  url = {https://link.aps.org/doi/10.1103/PhysRevB.99.174515}
}

@article{PhysRevB.92.174525,
  title = {Neutron scattering study of spin ordering and stripe pinning in superconducting ${\mathrm{La}}_{1.93}{\mathrm{Sr}}_{0.07}{\mathrm{CuO}}_{4}$},
  author = {Jacobsen, H. and Zaliznyak, I. A. and Savici, A. T. and Winn, B. L. and Chang, S. and H\"ucker, M. and Gu, G. D. and Tranquada, J. M.},
  journal = {Phys. Rev. B},
  volume = {92},
  issue = {17},
  pages = {174525},
  numpages = {13},
  year = {2015},
  month = {Nov},
  publisher = {American Physical Society},
  doi = {10.1103/PhysRevB.92.174525},
  url = {https://link.aps.org/doi/10.1103/PhysRevB.92.174525}
}

@article{PhysRevB.58.R2913,
  title = {Thermal conductivity of the spin-Peierls compound ${\mathrm{CuGeO}}_{3}$},
  author = {Ando, Yoichi and Takeya, J. and Sisson, D. L. and Doettinger, S. G. and Tanaka, I. and Feigelson, R. S. and Kapitulnik, A.},
  journal = {Phys. Rev. B},
  volume = {58},
  issue = {6},
  pages = {R2913--R2916},
  numpages = {0},
  year = {1998},
  month = {Aug},
  publisher = {American Physical Society},
  doi = {10.1103/PhysRevB.58.R2913},
  url = {https://link.aps.org/doi/10.1103/PhysRevB.58.R2913}
}

@article{PhysRevB.56.3402,
  title = {Electronic structure and magnetic properties of the linear chain cuprates ${\mathrm{Sr}}_{2}{\mathrm{CuO}}_{3}\phantom{\rule{0ex}{0ex}}$and ${\mathrm{Ca}}_{2}{\mathrm{CuO}}_{3}$},
  author = {Rosner, H. and Eschrig, H. and Hayn, R. and Drechsler, S.-L. and M\'alek, J.},
  journal = {Phys. Rev. B},
  volume = {56},
  issue = {6},
  pages = {3402--3412},
  numpages = {0},
  year = {1997},
  month = {Aug},
  publisher = {American Physical Society},
  doi = {10.1103/PhysRevB.56.3402},
  url = {https://link.aps.org/doi/10.1103/PhysRevB.56.3402}
}

@article{PhysRevB.107.214430,
  title = {Magnetic properties of $S=\frac{1}{2}$ distorted ${J}_{1}\ensuremath{-}{J}_{2}$ honeycomb lattice compound $\mathrm{NaCuIn}{({\mathrm{PO}}_{4})}_{2}$},
  author = {Singh, V. K. and Link, J. and Kargeti, K. and Barik, M. and Lenz, B. and Saraswat, N. and Jena, U. and Heinmaa, I. and Khuntia, P. and Boya, K. and Panda, S. K. and Stern, R. and Bitla, Y. and Chakrabarty, T. and Koteswararao, B.},
  journal = {Phys. Rev. B},
  volume = {107},
  issue = {21},
  pages = {214430},
  numpages = {11},
  year = {2023},
  month = {Jun},
  publisher = {American Physical Society},
  doi = {10.1103/PhysRevB.107.214430},
  url = {https://link.aps.org/doi/10.1103/PhysRevB.107.214430}
}

@article{PhysRevB.89.174403,
  title = {Microscopic magnetic modeling for the $S\phantom{\rule{0.16em}{0ex}}=\phantom{\rule{0.16em}{0ex}}\frac{1}{2}$ alternating-chain compounds ${\mathrm{Na}}_{3}{\mathrm{Cu}}_{2}{\mathrm{SbO}}_{6}$ and Na${}_{2}$Cu${}_{2}$TeO${}_{6}$},
  author = {Schmitt, M. and Janson, O. and Golbs, S. and Schmidt, M. and Schnelle, W. and Richter, J. and Rosner, H.},
  journal = {Phys. Rev. B},
  volume = {89},
  issue = {17},
  pages = {174403},
  numpages = {10},
  year = {2014},
  month = {May},
  publisher = {American Physical Society},
  doi = {10.1103/PhysRevB.89.174403},
  url = {https://link.aps.org/doi/10.1103/PhysRevB.89.174403}
}

@article{PhysRevB.91.144406,
  title = {Valence-bond solid as the quantum ground state in honeycomb layered urusovite $\mathrm{CuAl}(\mathrm{As}{\mathrm{O}}_{4})\mathrm{O}$},
  author = {Vasiliev, A. N. and Volkova, O. S. and Zvereva, E. A. and Koshelev, A. V. and Urusov, V. S. and Chareev, D. A. and Petkov, V. I. and Sukhanov, M. V. and Rahaman, B. and Saha-Dasgupta, T.},
  journal = {Phys. Rev. B},
  volume = {91},
  issue = {14},
  pages = {144406},
  numpages = {9},
  year = {2015},
  month = {Apr},
  publisher = {American Physical Society},
  doi = {10.1103/PhysRevB.91.144406},
  url = {https://link.aps.org/doi/10.1103/PhysRevB.91.144406}
}

@article{PhysRevLett.123.027201,
  title = {Novel Strongly Spin-Orbit Coupled Quantum Dimer Magnet: ${\mathrm{Yb}}_{2}{\mathrm{Si}}_{2}{\mathrm{O}}_{7}$},
  author = {Hester, Gavin and Nair, H. S. and Reeder, T. and Yahne, D. R. and DeLazzer, T. N. and Berges, L. and Ziat, D. and Neilson, J. R. and Aczel, A. A. and Sala, G. and Quilliam, J. A. and Ross, K. A.},
  journal = {Phys. Rev. Lett.},
  volume = {123},
  issue = {2},
  pages = {027201},
  numpages = {6},
  year = {2019},
  month = {Jul},
  publisher = {American Physical Society},
  doi = {10.1103/PhysRevLett.123.027201},
  url = {https://link.aps.org/doi/10.1103/PhysRevLett.123.027201}
}

@article{vasp1,
  title = {Ab initio molecular dynamics for liquid metals},
  author = {Kresse, G. and Hafner, J.},
  journal = {Phys. Rev. B},
  volume = {47},
  issue = {1},
  pages = {558--561},
  year = {1993},
  month = {Jan},
  publisher = {APS},
  url = {https://doi.org/10.1103/PhysRevB.47.558}
}

@article{vasp2,
  title = {Efficient iterative schemes for ab initio total-energy calculations using a plane-wave basis set},
  author = {Kresse, G. and Furthm\"uller, J.},
  journal = {Phys. Rev. B},
  volume = {54},
  issue = {16},
  pages = {11169--11186},
  year = {1996},
  month = {Oct},
  publisher = {APS},
  url = {https://doi.org/10.1103/PhysRevB.54.11169}
}

@article{gga,
  title = {Generalized Gradient Approximation Made Simple},
  author = {Perdew, J. P. and Burke, K. and Ernzerhof, M.},
  journal = {Phys. Rev. Lett.},
  volume = {77},
  issue = {18},
  pages = {3865--3868},
  year = {1996},
  month = {Oct},
  publisher = {APS},
  url = {https://doi.org/10.1103/PhysRevLett.77.3865}
}

@article{Cu-U1,
  title = {Band theory and Mott insulators: Hubbard U instead of Stoner I},
  author = {Anisimov, Vladimir I. and Zaanen, Jan and Andersen, Ole K.},
  journal = {Phys. Rev. B},
  volume = {44},
  issue = {3},
  pages = {943--954},
  year = {1991},
  month = {Jul},
  publisher = {American Physical Society},
  url = {https://doi.org/10.1103/PhysRevB.44.943}
}

@article{Cu-U2,
  title = {Pressure dependence of dynamically screened Coulomb interactions in NiO: Effective Hubbard, Hund, intershell, and intersite components},
  author = {Panda, S. K. and Jiang, H. and Biermann, S.},
  journal = {Phys. Rev. B},
  volume = {96},
  issue = {4},
  pages = {045137},
  year = {2017},
  month = {Jul},
  publisher = {American Physical Society},
  url = {https://doi.org/10.1103/PhysRevB.96.045137?_gl=1*hcutq5*_gcl_au*MTQ4MjYwMTAyMC4xNzM0Njg2MTI2*_ga*MTg4MzUyNTI3MC4xNzM0Njg2MTI1*_ga_ZS5V2B2DR1*MTczNjgzMzg5NS43LjEuMTczNjg0MTIxNy42MC4wLjEwNDIzNDM4NDE.}
}

@article{Cu-U3,
  title = {Effect of the electronic charge gap on LO bond-stretching phonons in undoped {L}a$_2${C}u{O}$_4$ calculated using $\rm{LDA}+\rm{U}$},
  author = {Sterling, Tyler C. and Reznik, Dmitry},
  journal = {Phys. Rev. B},
  volume = {104},
  issue = {13},
  pages = {134311},
  year = {2021},
  month = {Oct},
  publisher = {American Physical Society},
  url = {https://doi.org/10.1103/PhysRevB.104.134311}
}

@article{mft2,
  author = {Katsnelson, M. I. and Lichtenstein, A. I.},
  title = {First-principles calculations of magnetic interactions in correlated systems},
  journal = {Phys. Rev. B},
  volume = {61},
  issue = {13},
  pages = {8906--8912},
  year = {2000},
  month = {Apr},
  publisher = {American Physical Society},
  url = {https://doi.org/10.1103/PhysRevB.61.8906}
}

@article{mft1,
  title = {Local spin density functional approach to the theory of exchange interactions in ferromagnetic metals and alloys},
  author = {A.I. Liechtenstein and M.I. Katsnelson and V.P. Antropov and V.A. Gubanov},
  journal = {Journal of Magnetism and Magnetic Materials},
  volume = {67},
  issue = {1},
  pages = {65 - 74},
  year = {1987},
  publisher = {Elsevier},
  url = {https://doi.org/10.1016/0304-8853(87)90721-9}
}

@article{Bauer_2011,
doi = {10.1088/1742-5468/2011/05/P05001},
url = {https://doi.org/10.1088/1742-5468/2011/05/P05001},
year = {2011},
month = {may},
publisher = {},
volume = {2011},
number = {05},
pages = {P05001},
author = {Bauer, B and Carr, L D and Evertz, H G and Feiguin, A and Freire, J and Fuchs, S and Gamper, L and Gukelberger, J and Gull, E and Guertler, S and Hehn, A and Igarashi, R and Isakov, S V and Koop, D and Ma, P N and Mates, P and Matsuo, H and Parcollet, O and Pawłowski, G and Picon, J D and Pollet, L and Santos, E and Scarola, V W and Schollwöck, U and Silva, C and Surer, B and Todo, S and Trebst, S and Troyer, M and Wall, M L and Werner, P and Wessel, S},
title = {The ALPS project release 2.0: open source software for strongly correlated systems},
journal = {Journal of Statistical Mechanics: Theory and Experiment}
}

@article{SG1,
title = {Magnetic and electronic structure studies on S = 1/2 Heisenberg antiferromagnetic alternating spin chain system KCuGa(PO4)2},
journal = {Journal of Magnetism and Magnetic Materials},
volume = {630},
pages = {173391},
year = {2025},
issn = {0304-8853},
doi = {https://doi.org/10.1016/j.jmmm.2025.173391},
url = {https://www.sciencedirect.com/science/article/pii/S0304885325006237},
author = {V.K. Singh and S. Gayen and Sk. Soyeb Ali and R.C. Moharana and T. Chakrabarty and B. Koteswararao and S. Chattopadhyay and S.K. Panda}
}

@article{rodriguez1993recent,
  title={Recent advances in magnetic structure determination by neutron powder diffraction},
  author={Rodr{\'\i}guez Carvajal, Juan},
  journal={Physica B: Condens. Matter},
  volume={192},
  number={1-2},
  pages={55--69},
  year={1993},
  publisher={Elsevier}
}

@article{badri2023synthesis,
  title={Synthesis, crystals structures and theoretical studies of novel alkali copper indium orthophosphates ACuIn (PO4) 2 (A= Na, K and Rb)},
  author={Badri, Abdessalem and Slimi, Sami and Mokni, Ines and Dege, Necmi and Sol{\'e}, Rosa Maria and Aguil{\'o}, Magdalena and D{\'\i}az, Francesc and Mateos, Xavier and Amara, Mongi Ben},
  journal={Journal of Solid State Chemistry},
  volume={318},
  pages={123757},
  year={2023},
  publisher={Elsevier}
}

@book{selwood2013magnetochemistry,
  title={Magnetochemistry},
  author={Selwood, Pierce W},
  year={2013},
  publisher={Read Books Ltd}
}

@article{PhysRevB.89.014407,
  title = {Hindered magnetic order from mixed dimensionalities in ${\text{CuP}}_{2}{\text{O}}_{6}$},
  author = {Nath, R. and Ranjith, K. M. and Sichelschmidt, J. and Baenitz, M. and Skourski, Y. and Alet, F. and Rousochatzakis, I. and Tsirlin, A. A.},
  journal = {Phys. Rev. B},
  volume = {89},
  issue = {1},
  pages = {014407},
  numpages = {11},
  year = {2014},
  month = {Jan},
  publisher = {American Physical Society},
  doi = {10.1103/PhysRevB.89.014407},
  url = {https://link.aps.org/doi/10.1103/PhysRevB.89.014407}
}

@article{TM1,
  title = {Electronic structure and exchange interactions of insulating double perovskite ${\mathrm{La}}_{2}{\mathrm{CuRuO}}_{6}$},
  author = {Panda, S. K. and Kvashnin, Y. O. and Sanyal, B. and Dasgupta, I. and Eriksson, O.},
  journal = {Phys. Rev. B},
  volume = {94},
  issue = {6},
  pages = {064427},
  numpages = {8},
  year = {2016},
  month = {Aug},
  publisher = {American Physical Society},
  doi = {10.1103/PhysRevB.94.064427},
  url = {https://link.aps.org/doi/10.1103/PhysRevB.94.064427}
}

@article{TM2,
  title = {Strain-induced electronic and magnetic transition in the $S=\frac{3}{2}$ antiferromagnetic spin chain compound ${\mathrm{LaCrS}}_{3}$},
  author = {Kargeti, Kuldeep and Sen, Aadit and Panda, S. K.},
  journal = {Phys. Rev. B},
  volume = {109},
  issue = {3},
  pages = {035125},
  numpages = {9},
  year = {2024},
  month = {Jan},
  publisher = {American Physical Society},
  doi = {10.1103/PhysRevB.109.035125},
  url = {https://link.aps.org/doi/10.1103/PhysRevB.109.035125}
}

@article{TM3,
  title = {$S=1$ dimer system ${\mathrm{K}}_{2}\mathrm{Ni}{({\mathrm{MoO}}_{4})}_{2}$: A candidate for magnon Bose-Einstein condensation},
  author = {Lenz, B. and Koteswararao, B. and Biermann, S. and Khuntia, P. and Baenitz, M. and Panda, S. K.},
  journal = {Phys. Rev. B},
  volume = {106},
  issue = {18},
  pages = {L180408},
  numpages = {7},
  year = {2022},
  month = {Nov},
  publisher = {American Physical Society},
  doi = {10.1103/PhysRevB.106.L180408},
  url = {https://link.aps.org/doi/10.1103/PhysRevB.106.L180408}
}

@article{1D,
title = {Characterization of quasi-one-dimensional S=1/2 Heisenberg antiferromagnets Sr2Cu(PO4)2 and Ba2Cu(PO4)2 with magnetic susceptibility, specific heat, and thermal analysis},
journal = {Journal of Solid State Chemistry},
volume = {177},
number = {3},
pages = {883-888},
year = {2004},
issn = {0022-4596},
doi = {https://doi.org/10.1016/j.jssc.2003.09.024},
url = {https://www.sciencedirect.com/science/article/pii/S002245960300536X},
author = {A.A Belik and M Azuma and M Takano}
}

@article{spin-Peierls,
title = {Spin-flop and spin-Peierls transition in doped CuGeO3},
journal = {Spectrochimica Acta Part A: Molecular and Biomolecular Spectroscopy},
volume = {66},
number = {2},
pages = {307-311},
year = {2007},
issn = {1386-1425},
doi = {https://doi.org/10.1016/j.saa.2006.02.057},
url = {https://www.sciencedirect.com/science/article/pii/S1386142506001648},
author = {O. Yalçın and F. Yıldız and B. Aktaş}
}

@article{PAW,
  title = {Projector augmented-wave method},
  author = {Bl\"ochl, P. E.},
  journal = {Phys. Rev. B},
  volume = {50},
  issue = {24},
  pages = {17953--17979},
  numpages = {0},
  year = {1994},
  month = {Dec},
  publisher = {American Physical Society},
  doi = {10.1103/PhysRevB.50.17953},
  url = {https://link.aps.org/doi/10.1103/PhysRevB.50.17953}
}

@article{DFT1,
  title = {Self-Consistent Equations Including Exchange and Correlation Effects},
  author = {Kohn, W. and Sham, L. J.},
  journal = {Phys. Rev.},
  volume = {140},
  issue = {4A},
  pages = {A1133--A1138},
  numpages = {0},
  year = {1965},
  month = {Nov},
  publisher = {American Physical Society},
  doi = {10.1103/PhysRev.140.A1133},
  url = {https://link.aps.org/doi/10.1103/PhysRev.140.A1133}
}

@article{DFT2,
  title = {Inhomogeneous Electron Gas},
  author = {Hohenberg, P. and Kohn, W.},
  journal = {Phys. Rev.},
  volume = {136},
  issue = {3B},
  pages = {B864--B871},
  numpages = {0},
  year = {1964},
  month = {Nov},
  publisher = {American Physical Society},
  doi = {10.1103/PhysRev.136.B864},
  url = {https://link.aps.org/doi/10.1103/PhysRev.136.B864}
}

@article{rspt1,
  title = {Synthesis of band and model Hamiltonian theory for hybridizing cerium systems},
  author = {Wills, John M. and Cooper, Bernard R.},
  journal = {Phys. Rev. B},
  volume = {36},
  issue = {7},
  pages = {3809--3823},
  numpages = {0},
  year = {1987},
  month = {Sep},
  publisher = {American Physical Society},
  doi = {10.1103/PhysRevB.36.3809},
  url = {https://link.aps.org/doi/10.1103/PhysRevB.36.3809}
}

@article{GEK,
author = {Weiss, Alarich},
title = {John B. Goodenough: Magnetism and the Chemical Bond. Interscience Publishers. New York, London 1963. 393 Seiten, 89 Abbildungen. Preis: DM 95 s.},
journal = {Berichte der Bunsengesellschaft für physikalische Chemie},
volume = {68},
number = {10},
pages = {996-996},
doi = {https://doi.org/10.1002/bbpc.19640681015},
url = {https://onlinelibrary.wiley.com/doi/abs/10.1002/bbpc.19640681015},
eprint = {https://onlinelibrary.wiley.com/doi/pdf/10.1002/bbpc.19640681015},
year = {1964}
}

@misc{sandvik2019stochasticseriesexpansionmethods,
      title={Stochastic Series Expansion Methods}, 
      author={Anders W. Sandvik},
      year={2019},
      eprint={1909.10591},
      archivePrefix={arXiv},
      primaryClass={cond-mat.str-el},
      url={https://arxiv.org/abs/1909.10591}, 
}

@article{PhysRevLett.78.4063,
  title = {First-Principles Determination of the Soft Mode in Cubic ${\mathrm{ZrO}}_{2}$},
  author = {Parlinski, K. and Li, Z. Q. and Kawazoe, Y.},
  journal = {Phys. Rev. Lett.},
  volume = {78},
  issue = {21},
  pages = {4063--4066},
  numpages = {0},
  year = {1997},
  month = {May},
  publisher = {American Physical Society},
  doi = {10.1103/PhysRevLett.78.4063},
  url = {https://link.aps.org/doi/10.1103/PhysRevLett.78.4063}
}
\end{document}